\documentclass[aps,prb,reprint,twocolumn,superscriptaddress,showpacs,floatfix]{revtex4-1}

\usepackage{hyperref}
\usepackage{graphicx}
\usepackage{amsmath}
\usepackage{bbm}
\usepackage{color}
\usepackage{txfonts}
\usepackage{mathrsfs}
\usepackage{dcolumn}
\usepackage{multirow}
\usepackage{graphicx}
\usepackage{rotating}
\usepackage{layout}

\begin{document}

\title{Tunable cloaking of Mexican-hat confined states in bilayer silicene}
\date{\today}
\author{Le Bin Ho}
\email{binho@kindai.ac.jp}
\affiliation{Ho Chi Minh City Institute of Physics, VAST, Ho Chi Minh City 700000, Vietnam}
\affiliation{Department of Physics, Kindai University, Higashi-Osaka 577-8502, Japan}  
\author{Lan Nguyen Tran}
\email{lantrann@gmail.com}
\affiliation{Ho Chi Minh City Institute of Physics, VAST, Ho Chi Minh City 700000, Vietnam}

\begin{abstract}
 We present the ballistic quantum transport of a {\it p-n-p} bilayer silicene junction in the presence of spin-orbit coupling and electric field using a four-band model in combination with the transfer-matrix approach. A Mexican-hat shape of low-energy spectrum is observed similarly to bilayer graphene under an interalayer bias. We show that while bilayer silicene shares some physics with bilayer graphene, it has many intriguing phenomena that have not been reported for the latter. First, the confined state producing a significantly non-zero transmission in Mexican hat. Second, the cloaking of the Mexican-hat confined state is found. Third, we observe that the Mexican-hat cloaking results in a strong oscillation of conductance when the incident energy is below the potential height. Finally, unlike monolayer silicene, the conductance at large interlayer distances increases with the rise of electric field when the incident energy is above the potential height.  
\end{abstract}
\maketitle

        Unlike monolayer graphene, bilayer graphene has a parabolic dispersion relation and no Klein tunneling is observed \cite{mccann2013electronic,katsnelson2006chiral}. The chirality mismatch of states inside and outside a {\it p-n-p} junction leads to a cloaking of transmission in a certain region of incident energy\cite{gu2011chirality,van2013four}. More interestingly, applying different electrostatic potentials at the two layers of bilayer graphene, called biased bilayer graphene, results in a tunable band gap and Mexican-hat shape of low-energy spectrum \cite{castro2010electronic}. Great efforts both in theory and experiment have been devoted to reproduce and explain these phenomena \cite{lee2016evidence,mak2009observation,castro2010electronic}. Thanks to its peculiar electronic structures, biased bilayer graphene was proposed as a new platform for electronic devices, such as the low-voltage tunnel switches \cite{alymov2016abrupt}. Moreover, some recent studies have revealed a hydrogen-like bound state within Mexican hat opening a new door for biased bilayer graphene applications \cite{skinner2014bound}.

While sharing some intriguing properties of graphene, silicene, a two-dimensional allotrope of silicon, has some superior advantages compared to graphene, such as strong spin-orbit coupling (SOC) and buckled honeycomb structure.
While SOC enables us to realize the quantum spin Hall effect \cite{Liu2011}, the buckled honeycomb structure help us control the bulk band gap of silicene by applying an external electric field \cite{Drummond2012}.
Topological phase transitions and quantum transport properties of monolayer silicene in the presence of external fields, such as electric and magnetic exchange fields, and circularly polarized light in the off-resonant regime, have been extensively reported \cite{Ezawa_prl2012,Ezawa_prl2013,ho2016photoenhanced}.

Apart from monolayer, bilayer silicene were also successfully synthesized in experiment. It is expected that bilayer silicene can provide some unusal physics that cannot be found in monolayer. Recently, there have been many theoretical works focusing on the topological phase transitions, magneto-optical, and optoelectronic properties of bilayer silicene, for instances, see Refs.~[\cite{ezawa2012quasi,da2017magneto,huang2014exceptional}].
Its balistic transport properties, however, have not been widely investigated. As seen from bilayer graphene, the two-band model is in sufficient in the presence of a strong interlayer bias even at the Dirac point \cite{castro2010electronic,van2013four}. Therefore, a four-band model is essential in order to properly describe the low-energy physics of bilayer silicene.

In this paper, we investigate ballistic transport properties of a {\it p-n-p} bilayer silicene junction in the presence of a transverse electric field using a four-band low-energy model. The transfer-matrix approach was implemented to evaluate transmission spectra. Some novel physics that have not been reported for monolayer silicene and bilayer graphene were observed. We found that there is a non-zero transmission within the Mexican hat, indicating confined states within this region. Moreover, the cloaking of these states results in a strong oscillation of conductance with respect to electric field when the incident energy is below the potential height. On the other hand, unlike monolayer, the conductance of bilayer silicene is enhanced under electric field when the incident energy is above the potential height.

\begin{figure}[h]
  \includegraphics[width=6cm,]{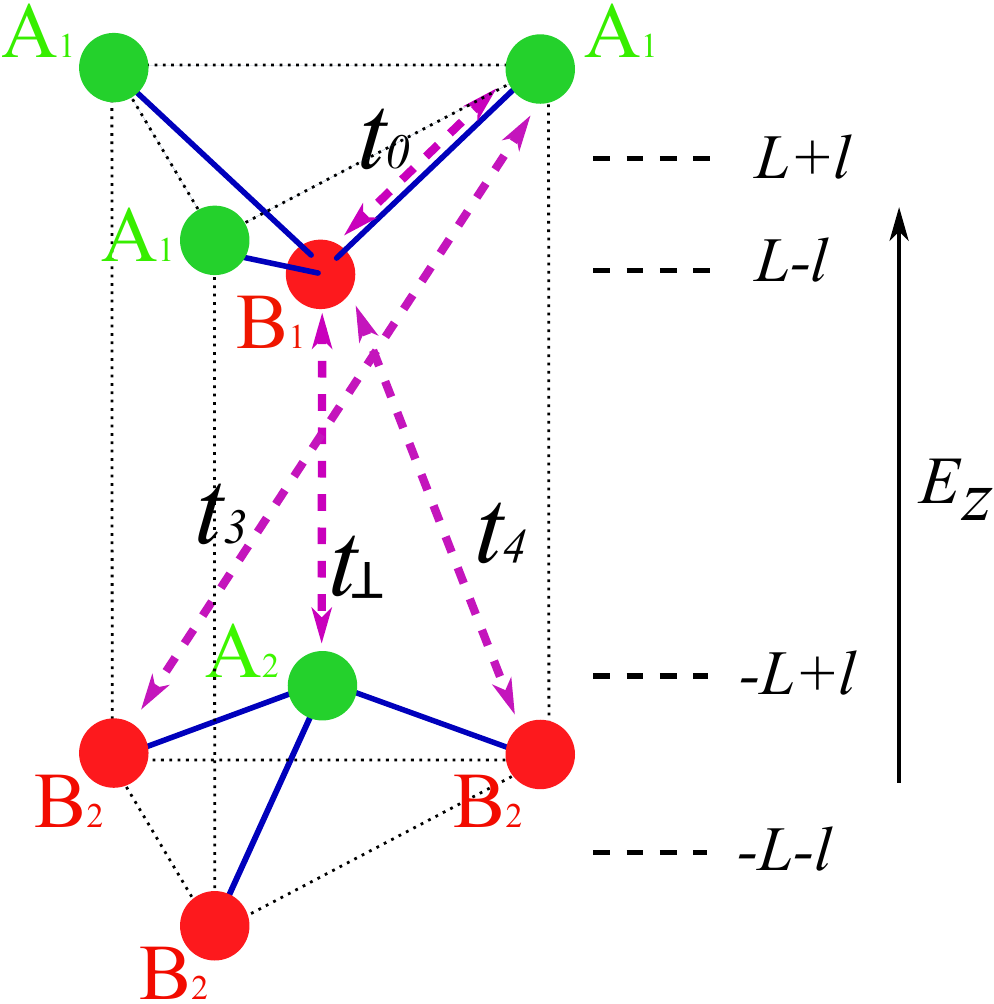}
  \caption{ An unit cell of AB bilayer silicene with a forward stacking configuration. Green and red indicate the two sublattices A and B of monolayer, respectively. $L$ is the interlayer distance, while 2$l$ is intralayer sublattice distance. $t_0$ is the intralayer hoping, $t_{\perp}$ is the perpendicular interlayer hoping, $t_3$ and $t_4$ are interlayer skew hoping that are not included in the current work. 
  }
  \label{fig:geom}
\end{figure}

While there are four possibilities of $AB$ bilayer stacking\cite{ezawa2012quasi}, we only consider the forward stacking configuration displayed in Figure~\ref{fig:geom}. As seen in the figure, bilayer silicene are composed of two silicene monolayers having an in-plane interatomic distance $a = 2.46\r{A}$. Each layer has a buckled structure consisting of two nonequivalent sublattices denoted by $A$ and $B$. The intralayer atomic distance is 2$l$ with $l=0.23\AA$. The spin-orbit coupling $\lambda_{SO}$ and the intralayer coupling between $A$ and $B$ atoms $t_0$ are $3.9$meV and $1.6$eV, respectively. 
The two layers are stacked according to a Bernal $A_1B_2$ stacking where $A_1$ is right above $B_2$ with a distance $L$. 
As shown in Figure~\ref{fig:geom}, the perpendicular interlayer coupling between the $A_2$ and $B_1$ atoms is $t_{A_2B_1}=t_{\perp}$, while those between the other interlayer atom pairs are $t_{A_1B_2}=t_{A_2B_1}=t_3$ and $t_{A_1A_2}=t_{B_1B_2}=t_4$. The interlayer skew hopping term $t_3$ results in a so-called trigonal warping occurring only at very low energies. The second skew hopping term, $t_4$, has a tiny impact on the electronic properties. Therefore, we have not included these two in the current work.

Following the continuum nearest-neighbor tight-binding formalism, the effective Hamiltonian near the Dirac points and the eigenstate are given by \cite{ezawa2012quasi,van2013four,castro2010electronic}
\small
\begin{align}{\label{H0}}
H = 
 \begin{pmatrix}
 U+m_+ & v_F\pi & t_{\perp} & 0 \\
  v_F\pi^\dagger & U+m_- & 0 & 0 \\
  t_{\perp} & 0  & U-m_+ & v_F\pi^\dagger  \\
  0 & 0 &v_F\pi & U-m_- 
 \end{pmatrix},
 \hspace {0.1cm}
 \Psi =  \begin{pmatrix}
  \psi_{A_1} \\
 \psi_{B_1} \\
  \psi_{B_2}  \\
  \psi_{A_2} 
 \end{pmatrix},
\end{align}
\normalsize
where $v_F \approx 5.5\times 10^5$ m/s is the Fermi velocity of the charge carries in silicene, $\pi=p_x+ip_y$ and $\bm{p}$ is the momentum operator, U is an external potential. 
The terms $m_\pm$ represent the contribution of SOC ($\lambda_{SO}$) and electric field $E_z$. For the forward stacking configuration, we have  $m_\pm = \mp\lambda_{SO}+(L\pm l)E_z$. Eigenvalues of the Hamiltonian~\ref{H0} are then given by
\begin{align}{\label{spec}}
\epsilon = \eta \dfrac{1}{\sqrt{2}}\sqrt{\beta+ \theta \sqrt{\beta^2-4\alpha}},
\end{align}
where the index $\eta = \pm1$ corresponds to conducting (+) and valence (--) bands , while the index $\theta = \pm1$ represents the low-energy (--) and high-energy (+) branches. $\beta = 1+m_+^2+m_-^2+2k^2$ with $k = \sqrt{k_x^2 + k_y^2}$ and $\alpha = (k^2-m_+m_-)^2+m_-^2$. 
We have used the dimensionless variable, $\epsilon = (E-V)/t_{\perp}$, $k_y \to \hbar v_Fk_y /t_{\perp}$. 
In the presence of SOC and electric field, the low-energy branches ($\theta=-1$) of band structure (\ref{spec}) displays an unique Mexican-hat shape as seen in the upper panel of Figure~\ref{fig:bands}. 
In the limit $\epsilon \ll t_{\perp}$ and with an assumption that $m_{\pm}$ and $\epsilon$ are the same order of magnitude, one can obtain two-band approximation \cite{van2013four,castro2010electronic}. Clearly, the two-band model is unable to yield the Mexican-hat shape as seen in the upper panel of Figure~{\ref{fig:bands}. We therefore will not discuss it further in this paper.

The lower panel of Figure~\ref{fig:bands} represents the variation of bilayer silicene band gap as electric field $E_z$ increases. For comparison, the monolayer result is also provided. Critical points where the band gap is closed are observed for both systems, however, it is lower for the bilayer than for the monolayer one. More importantly, while the band gap linearly increases beyond the critical point for monolayer, it is slowly enlarged and almost unchanged when $E_z$ is larger than 1.0$t_\perp$ for bilayer.    

\begin{figure}[h!]
  \includegraphics[width=5cm]{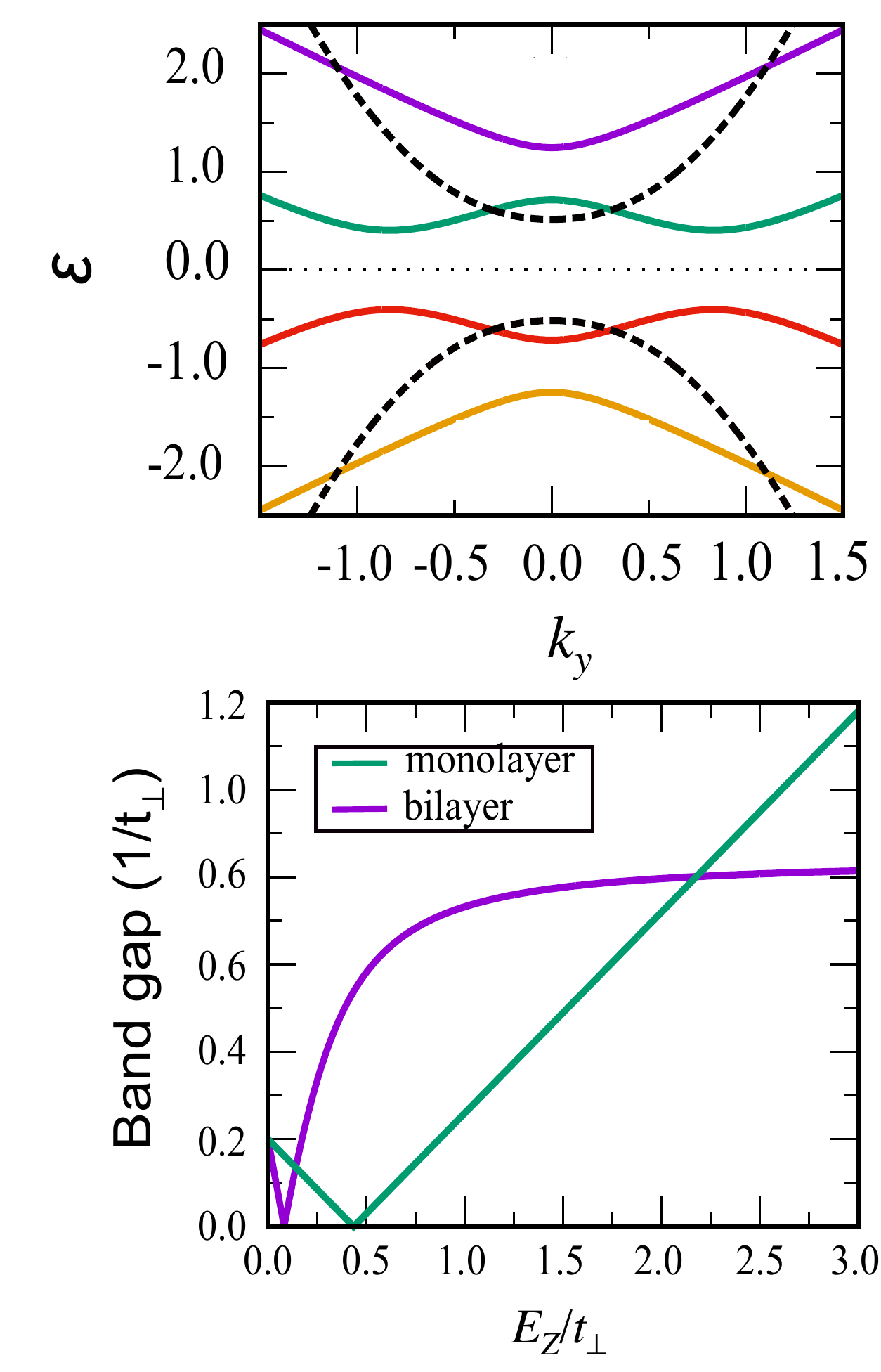}
  \caption{ Upper panel: band structures of bilayer silicene with $\lambda_{SO} = 0.1t_{\perp}$, $E_z = 0.5t_{\perp}$, and $U = 0$. The dot black curve correspond the spectrum of two-band approximation. Lower panel: the band gap of bilayer silicene as functions of electric field $E_z$. The monolayer result is also plotted for comparison.}
  \label{fig:bands}
\end{figure}

We now model a one-dimensional square well potential $V(x)$ of a width $d$ applied equally to the two layers of bilayer silicene as follows 
\begin{align}{\label{potential}}
V(x) = \left\{ \begin{array}{ll}
         V & \mbox{if $0 \leq x \leq d$ (region 2)};\\
         0 & \mbox{if $x < 0$ or $x > d$ (region 1 or 3)}.\end{array} \right.
\end{align}
The wave function can be written as $\Psi(x,y) = \psi(x)e^{ik_yy}$, with the translational invariance along the $y$ direction, i.e. the momentum $k_y$ is unchanged during electron motion. The electric field is only applied to the region 2.

Solving the time-independent Schrodinger equation $H\Psi = E\Psi$ we obtain the eigenstates, that are given as
\begin{align}{\label{wave}}
\psi(x) = 
 \begin{pmatrix}
\psi_{A_1} \\
 \psi_{B_1}  \\
 \psi_{B_2}  \\
 \psi_{A_2} 
 \end{pmatrix}
 = \mathcal{P}\mathcal{Q}(x)
C,
 \end{align}
Here, $C$ are wavefunction coefficients, $\mathcal{Q}(x) = {\rm diag} (e^{ik_+x},e^{-ik_+x},e^{ik_-x},e^{-ik_-x})$ and
\begin{align}{\label{P}}
\mathcal{P} = 
 \begin{pmatrix}
  1 & 1 & 1 & 1 \\
  f^+_+ & f^+_- & f^-_+ & f^-_- \\
   g^+ & g^+ & g^- & g^- \\
   h^+_+ & h^+_- & h^-_+ & h^-_- 
 \end{pmatrix}.
 \end{align}
with,  
\begin{align}
 f^{\eta}_{\pm} &= (\pm k_{\eta}-ik_y)/(E-m_-), \nonumber \\ 
 g^{\eta} &= [-k_{\eta}^2-k_y^2+\gamma_{-}]/(E-m_-), \nonumber \\  
 h^{\eta}_\pm &= [-k_{\eta}^2-k_y^2+\gamma_{-}](\pm k_{\eta}+ik_y)/(E^2-m_-^2),\nonumber
\end{align}
where
\begin{align}\label{eq:k_eta}
  k_{\eta} &= \sqrt{\frac{\gamma_++\gamma_-}{2}+\eta\sqrt{\Delta} -k_y^2},
\end{align}  
and
\begin{align}  
  \gamma_{\pm} &= (E\pm m_{-})(E\pm m_{+}), \nonumber \\
  \Delta &= \frac{(\gamma_++\gamma_-)^2}{4}+(E^2-m_-^2). \nonumber 
\end{align}
$k_{\eta}$ is the wave vector in the $x$ direction, with $\eta = \pm 1$. It is derived from the dispersion relation (Eq.~\ref{spec}). The index $\eta$ now corresponds to the pseudospin state of particles. 
The wave vector $k_+$ is always real whenever $E\geq \lambda_{SO}$, which is the case we consider in this paper. The wave vector $k_-$, however, can be real or imaginary due to the relation of the value $E$ to $\lambda_{SO}$, and $k_y$. For the normal incident, $k_y = 0$, when $\lambda_{SO}<E<\sqrt{1+\lambda_{SO}^2}$, $k_-$ is imaginary, therefore, the propagation only happens for the $k_+$ mode. When $E>\sqrt{1+\lambda_{SO}^2}$, $k_-$ becomes real, the propagation is then carried out by both modes. Corresponding to the two distinct propagate modes, there are two nonscattering transmission channels as $T^+_+$ and $T^-_-$ for propagation via $k_+$ and $k_-$ modes, respectively. There also exists two others scattering channels $T^+_-$ for scattering from $k_+$ to $k_-$, and $T^-_+$ for scattering from $k_-$ to $k_+$.

The continuity of wave functions at $x=0$ and $x=d$ gives the boundary conditions $\psi_1(0)=\psi_2(0)$ and $\psi_2(d)=\psi_3(d)$. Note that the electric field is only applied to the region 2. The transfer matrix $\mathcal{M}$ can be then written as 
\begin{align}
    \mathcal{M}=\mathcal{P}_1^{-1}\mathcal{P}_2\mathcal{Q}_2^{-1}(d)\mathcal{P}_2^{-1}\mathcal{P}_3\mathcal{Q}_3(d),    
\end{align}
and the components of the vector $C$ in the region I and III are given:
\begin{align}{\label{c_vector}}
C^{\eta}_I = 
 \begin{pmatrix}
\delta_{l,1} \\
r^{\eta}_+ \\
 \delta_{l,-1}  \\
 r^{\eta}_- 
 \end{pmatrix},
 \hspace{0.05cm} \text {and} \hspace{0.2cm}
 C^{\eta}_{III} = 
 \begin{pmatrix}
t^{\eta}_+\\
0 \\
r^{\eta}_- \\
0
 \end{pmatrix},
 \end{align}
where $l=\pm 1.$ Finally, by taking into account the change in velocity of the waves scattering into different modes, the transmissions $T$ are obtained as
\begin{align}{\label{T}}
T_\pm^{\eta} = \dfrac{k_\pm}{k_{\eta}}|t^{\eta}_\pm|^2.
\end{align}

The normalized spin-valley dependent conductance at zero temperature is evaluated according to Landauer-B\"{u}ttiker formalism as
\begin{align}
  G = \frac{1}{2}\int_{-\pi/2}^{\pi/2} \sum T_\pm^\pm (E,\phi)\cos(\phi)d\phi,
\end{align}
where $\phi$ is the incident angle.

In unbiased bilayer graphene, the cloaking effect of transmission through a barrier was observed at normal incidence \cite{gu2011chirality,van2013four}. This can be briefly explained as follows. Let us consider a propagation via the $k_+$ mode as displayed in Figure~\ref{fig:cloaking}. For normal incidence ($k_y = 0$), the pseudospin is conserved. This means that the $k_+$ mode outside the barrier can only couple with the $k_+$ mode inside the barrier. However, the energy spectrum inside the barrier is shifted, leading to the mismatch between $k_+$ modes inside and outside the barrier. Even though there are $k_-$ states available inside the barrier, the propagation via the $k_+$ mode through the barrier is unlikely, resulting in the transmission cloaking of confined states inside the barrier. 

\begin{figure}[t]
  \includegraphics[width=7cm]{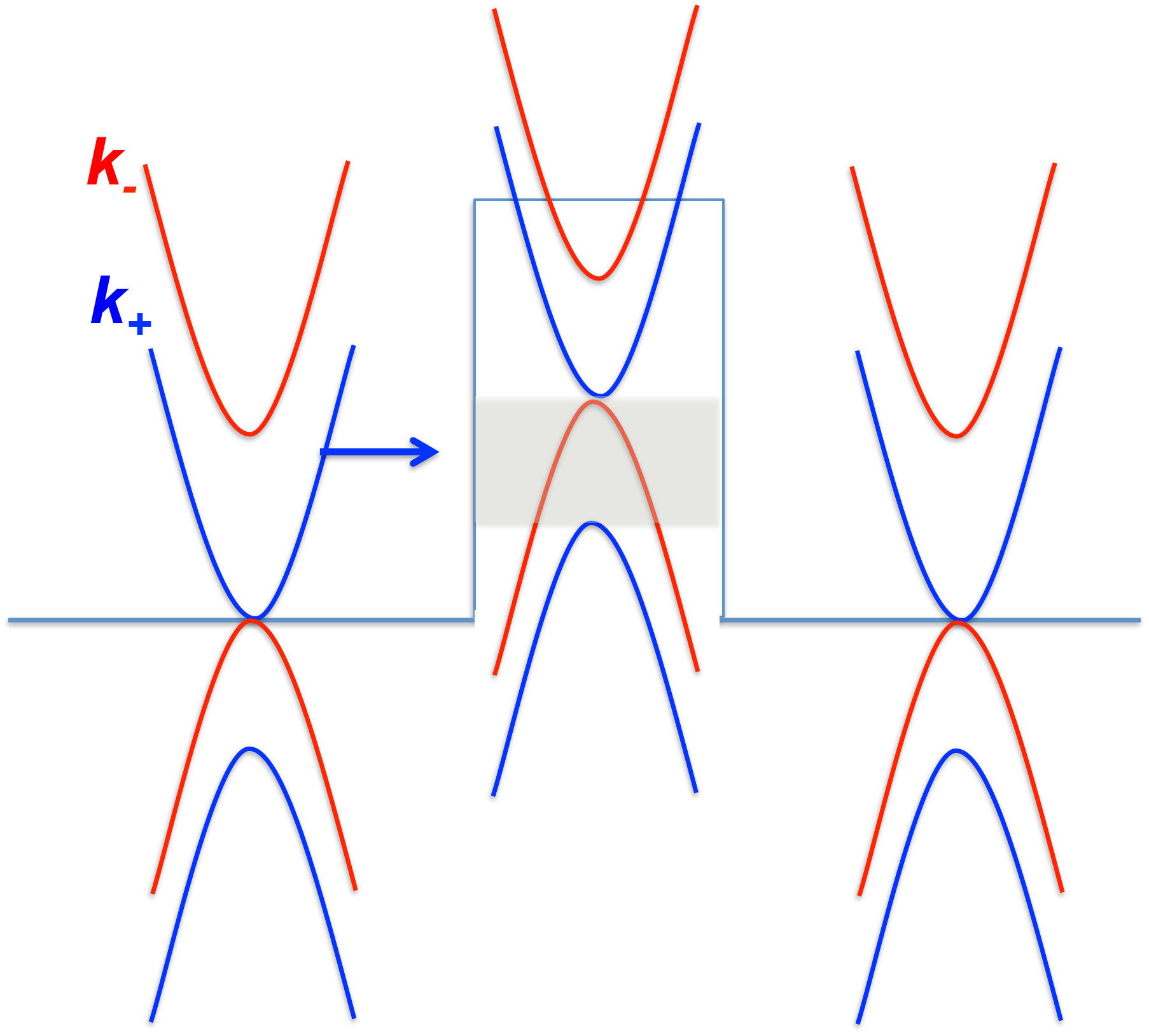}
  \caption{Schematic representation of energy spectra of unbised bilayer graphene inside and outside the potential barrier $U$. The arrow indicates the direction of propagation. The transmission cloaking of $k_+$ mode occurs in the gray region where there are no available $k_+$ states inside the barrier.}
   \label{fig:cloaking}
\end{figure}

\begin{figure*}[t!]
  \includegraphics[width=12cm]{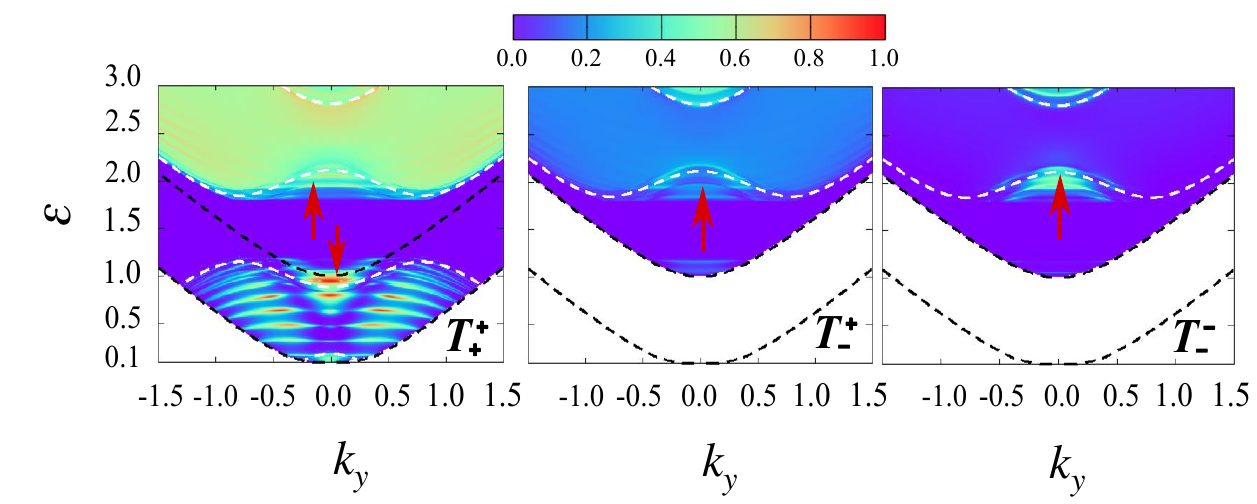}
  \caption{Transmission spectra of different modes ($T^+_+, T^+_-=T^-_+$, and $T^-_-$) as functions of incident energy and transverse wave vector $k_y$ in the presence of SOC $\lambda_{SO} = 0.1t_{\perp}$ and electric field $E_z = 0.5t_{\perp}$. The white dashed curves are the four-band dispersion spectra. Whereas, the black dashed curves are the border between the propagating or evanescent regions. The red arrows indicate the non-zero transmission within the Mexican hat. The height of potential barrier is 1.5$t_\perp$. The interlayer distance  $L = 1.5 \r{A}$.}
  \label{fig:T}
\end{figure*}

We now focus on the transmission spectra of bilayer silicene in the presence of SOC and electric field displayed in Figure~\ref{fig:T}. 
Since the electric field $E_z$ modifies the particles' momenta $k_{\eta}$ inside the barrier (Eq.~\ref{eq:k_eta}), the cloaking in the $T^+_+$ channel splits into two branches at finite $k_y$ instead of at normal incidence ($k_y = 0$). The splitting of the transmission cloaking was also found for bilayer graphene in the presence of interlayer bias \cite{van2013four}. 
One fascinating feature that was not reported for bilayer graphene is that transmission within the Mexican hat is significantly non-zero for all channels, as indicated by red arrows in the figure, implying the existence of confined states in this region. 

\begin{figure}[t]
  \includegraphics[width=6cm]{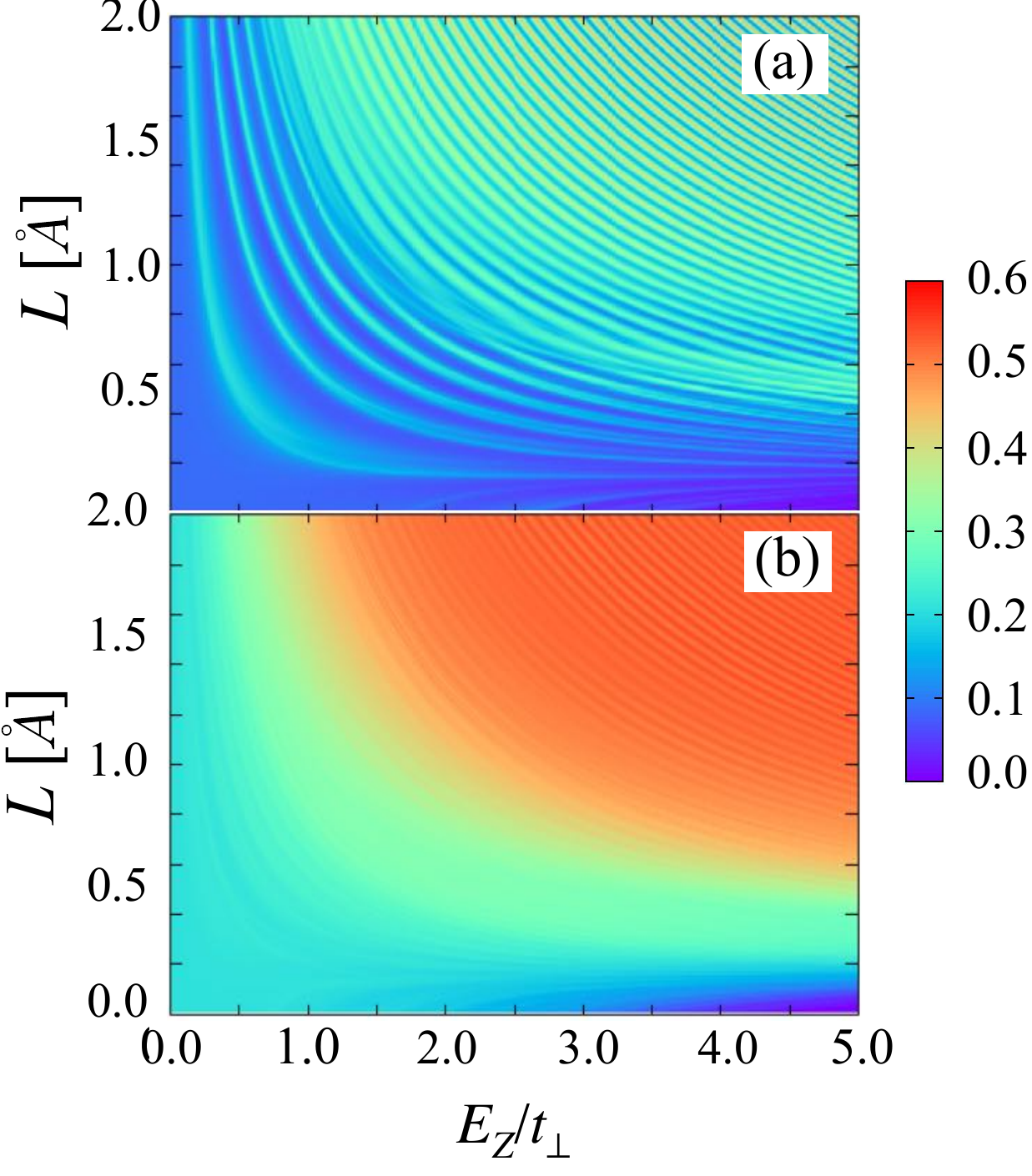}
  \caption{Conductance as a function of the distance between two layers $L$ and $E_z$ at different values of incident energy: (a) $E = 1.0t_{\perp}$ and (b) $E = 2.0t_{\perp}$. The potential height $U = 1.5t_{\perp}$.}
   \label{fig:conduc}
\end{figure}

Figure~\ref{fig:conduc} represents the conductance as a function of the interlayer distance $L$ and electric field $E_z$. The potential height $U$ is set at $1.5 t_{\perp}$. Two different values of the incident energy $E$ are considered: below ($ 1.0t_{\perp}$) and above ($2.0t_{\perp}$) potential height. Unlike monolayer silicene where the conductance monotonously decreases resulted from a linear dependence of band gap on electric field as seen in Figure~\ref{fig:bands}, some new phenomena are observed for bilayer silicene. 

At $E = 1.0t_{\perp}$ (Figure~\ref{fig:conduc}a), the conductance strongly oscillates with respect to both $L$ and $E_z$. In order to get insight into this behavior, we plot in Figure~\ref{fig:T++} the $T^+_+(E,k_y)$ spectrum at three selected values of electric field, $E_z =$ 0.9$t_{\perp}$, 0.98$t_{\perp}$, and 1.05$t_{\perp}$ associated with peaks and valley of conductance. The interlayer distance is fixed at 1.5 \r{A}. We only consider the channel $T^+_+$ because, as seen in Figure~\ref{fig:T}, this channel dominates the conductance when the incident energy below the potential height, e.g. valence band. As seen in Figure~\ref{fig:T++}, the $T^+_+$ transmission of valence band is mainly contributed by confined states in the Mexican hat. Strong Fabry-Perot resonances of transmission spectra imply the discretization of these states. Furthermore, the transmission within Mexican hat oscillates with respect to the wave vector $k_y$. At $E_z = 0.9t_{\perp}$, there is a large cloaking region around the normal incidence ($k_y = 0$) significantly suppressing the conductance. Whereas, at $E_z = 0.98t_{\perp}$, the cloaking shifts to finite $k_y$ and is less significant leading to an enhancement of conductance.  At $E_z = 1.05t_{\perp}$, the conductance is again lowered due to a large cloaking at normal incidence. In general, the cloaking at normal incidence within the Mexican hat causes the oscillation of conductance. 

{\it What is the origin of the transmission cloaking in the Mexican hat?} It is believed not due to a shift of energy spectrum as in the case of barrier potential discussed above. Recently, Skinner and coworkers \cite{skinner2014bound} have found a hydrogen-like bound state within Mexican hat of biased bilayer graphene. They showed that the bound state's electron density strongly oscillates with respect to the wave vector $k$ as the applied bias increases. Following this argument, we may attribute the cloaking of the confined state inside Mexican hat to the oscillation of its electron density. For example, at $E_z = 0.9t_{\perp}$, the confined state's electron density is almost zero around the normal incidence. Therefore, it does not show up in the normal incidence transmission. In contrast, the confined state's electron density around normal incidence is largely non-zero at $E_z = 0.98t_{\perp}$. As a result, the normal incidence propagation via Mexican hat is allowed.  

In order to more convincingly demonstrate the cloaking of Mexican-hat confined state, an analytic relation between its electron density and electric field is essentially derived. However, it is beyond the scope of the current work and we would leave it for a future work. 

\begin{figure*}[t]
  \includegraphics[width=12cm,height=5cm]{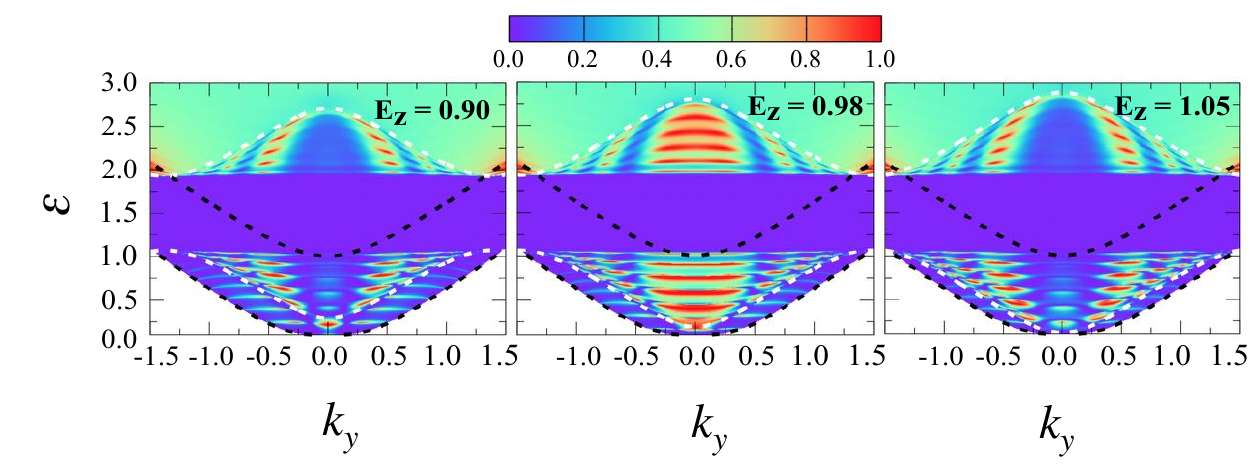}
  \caption{$T^+_+$ as a function of incident energy $E$ and transverse wave vector $k_y$ at different value of $E_z$. The white dashed curves are the four-band dispersion spectra. Whereas, the black dashed curves are the border between the propagating or evanescent regions. The red arrows indicate the non-zero transmission within the Mexican hat. The potential height $U = 1.5t_\perp$. The interlayer distance  $L = 1.5 \r{A}$.}
  \label{fig:T++}
\end{figure*}

The conductance at an incident energy above the potenital height, $E = 2.0t_{\perp}$, is presented in Figure~\ref{fig:conduc}b. Even though there is a large tranmission cloaking in the Mexican hat, the oscillation of conductance is less significant than it is at $E = 1.0t_{\perp}$. When the distance is small, $L < 0.2$ \r{A}, the conductance monotonously decreases with the increasing of electric field that is typical for monolayer silicene. As seen in Figure~\ref{fig:T++}, the transmission for the conducting band is contributed from both states inside and outside the Mexican hat. On the other hand, the band gap tends to a saturation when $E_z > 1.0t_{\perp}$ as seen in Figure~\ref{fig:bands}. As a consequence, unlike monolayer, enlarging the interlayer distance results in a slow increasing of conductance with the rise of electric field $E_z$.

In conclusions, we have presented the ballistic transport of a {\it p-n-p} bilayer silicene junction using a four-band model in the presence of both SOC and electric field. A Mexican-hat shape of low-energy spectra was observed similarly to biased bilayer graphene. It was shown that while bilayer silicene shares some properties with bilayer graphene, it has many intriguing phenomena that have not been reported for biased bilayer graphene. First, a non-zero transmission is observed within Mexican hat suggesting the existence of a localized state in this region. Second, for incident energy below the potential height, the conductance strongly oscillates when interlayer distance and electric field increase. This oscillation can be attributed the cloaking of the Mexican-hat confined state that is originated from the oscillation of its electron density with respect to wave vector. Third, for incident energy above the potential height, unlike monolayer, the conductance at the large interlayer distance increases with the rise of electric field. Our theoretical results are believed to be useful for realistic applications of bilayer silicene in electronics, such as field effect transistors or electronic switches. Working on analytic relation between the Mexican-hat confined state's electron density and electric field is on progress.

{\it Acknowledgment.} This work was supported by Vietnamese National Foundation of Science and Technology Development (NAFOSTED) under Grant No. 103.01-2015.14.

\bibliography{main}

\begin{thebibliography}{17}%
\makeatletter
\providecommand \@ifxundefined [1]{%
 \@ifx{#1\undefined}
}%
\providecommand \@ifnum [1]{%
 \ifnum #1\expandafter \@firstoftwo
 \else \expandafter \@secondoftwo
 \fi
}%
\providecommand \@ifx [1]{%
 \ifx #1\expandafter \@firstoftwo
 \else \expandafter \@secondoftwo
 \fi
}%
\providecommand \natexlab [1]{#1}%
\providecommand \enquote  [1]{``#1''}%
\providecommand \bibnamefont  [1]{#1}%
\providecommand \bibfnamefont [1]{#1}%
\providecommand \citenamefont [1]{#1}%
\providecommand \href@noop [0]{\@secondoftwo}%
\providecommand \href [0]{\begingroup \@sanitize@url \@href}%
\providecommand \@href[1]{\@@startlink{#1}\@@href}%
\providecommand \@@href[1]{\endgroup#1\@@endlink}%
\providecommand \@sanitize@url [0]{\catcode `\\12\catcode `\$12\catcode
  `\&12\catcode `\#12\catcode `\^12\catcode `\_12\catcode `\%12\relax}%
\providecommand \@@startlink[1]{}%
\providecommand \@@endlink[0]{}%
\providecommand \url  [0]{\begingroup\@sanitize@url \@url }%
\providecommand \@url [1]{\endgroup\@href {#1}{\urlprefix }}%
\providecommand \urlprefix  [0]{URL }%
\providecommand \Eprint [0]{\href }%
\providecommand \doibase [0]{http://dx.doi.org/}%
\providecommand \selectlanguage [0]{\@gobble}%
\providecommand \bibinfo  [0]{\@secondoftwo}%
\providecommand \bibfield  [0]{\@secondoftwo}%
\providecommand \translation [1]{[#1]}%
\providecommand \BibitemOpen [0]{}%
\providecommand \bibitemStop [0]{}%
\providecommand \bibitemNoStop [0]{.\EOS\space}%
\providecommand \EOS [0]{\spacefactor3000\relax}%
\providecommand \BibitemShut  [1]{\csname bibitem#1\endcsname}%
\let\auto@bib@innerbib\@empty
\bibitem [{\citenamefont {McCann}\ and\ \citenamefont
  {Koshino}(2013)}]{mccann2013electronic}%
  \BibitemOpen
  \bibfield  {author} {\bibinfo {author} {\bibfnamefont {E.}~\bibnamefont
  {McCann}}\ and\ \bibinfo {author} {\bibfnamefont {M.}~\bibnamefont
  {Koshino}},\ }\href {https://doi.org/10.1088/0034-4885/76/5/056503}
  {\bibfield  {journal} {\bibinfo  {journal} {Rep. Prog. Phys.}\ }\textbf
  {\bibinfo {volume} {76}},\ \bibinfo {pages} {056503} (\bibinfo {year}
  {2013})}\BibitemShut {NoStop}%
\bibitem [{\citenamefont {Katsnelson}\ \emph {et~al.}(2006)\citenamefont
  {Katsnelson}, \citenamefont {Novoselov},\ and\ \citenamefont
  {Geim}}]{katsnelson2006chiral}%
  \BibitemOpen
  \bibfield  {author} {\bibinfo {author} {\bibfnamefont {M.}~\bibnamefont
  {Katsnelson}}, \bibinfo {author} {\bibfnamefont {K.}~\bibnamefont
  {Novoselov}}, \ and\ \bibinfo {author} {\bibfnamefont {A.}~\bibnamefont
  {Geim}},\ }\href {https://doi.org/10.1038/nphys384} {\bibfield  {journal}
  {\bibinfo  {journal} {Nat. Phys.}\ }\textbf {\bibinfo {volume} {2}},\
  \bibinfo {pages} {620} (\bibinfo {year} {2006})}\BibitemShut {NoStop}%
\bibitem [{\citenamefont {Gu}\ \emph {et~al.}(2011)\citenamefont {Gu},
  \citenamefont {Rudner},\ and\ \citenamefont {Levitov}}]{gu2011chirality}%
  \BibitemOpen
  \bibfield  {author} {\bibinfo {author} {\bibfnamefont {N.}~\bibnamefont
  {Gu}}, \bibinfo {author} {\bibfnamefont {M.}~\bibnamefont {Rudner}}, \ and\
  \bibinfo {author} {\bibfnamefont {L.}~\bibnamefont {Levitov}},\ }\href
  {https://doi.org/10.1103/PhysRevLett.107.156603} {\bibfield  {journal}
  {\bibinfo  {journal} {Phys. Rev. Lett.}\ }\textbf {\bibinfo {volume} {107}},\
  \bibinfo {pages} {156603} (\bibinfo {year} {2011})}\BibitemShut {NoStop}%
\bibitem [{\citenamefont {Van~Duppen}\ and\ \citenamefont
  {Peeters}(2013)}]{van2013four}%
  \BibitemOpen
  \bibfield  {author} {\bibinfo {author} {\bibfnamefont {B.}~\bibnamefont
  {Van~Duppen}}\ and\ \bibinfo {author} {\bibfnamefont {F.}~\bibnamefont
  {Peeters}},\ }\href {https://doi.org/10.1103/PhysRevLett.107.156603}
  {\bibfield  {journal} {\bibinfo  {journal} {Phys. Rev. B}\ }\textbf {\bibinfo
  {volume} {87}},\ \bibinfo {pages} {205427} (\bibinfo {year}
  {2013})}\BibitemShut {NoStop}%
\bibitem [{\citenamefont {Castro}\ \emph {et~al.}(2010)\citenamefont {Castro},
  \citenamefont {Novoselov}, \citenamefont {Morozov}, \citenamefont {Peres},
  \citenamefont {Dos~Santos}, \citenamefont {Nilsson}, \citenamefont {Guinea},
  \citenamefont {Geim},\ and\ \citenamefont {Neto}}]{castro2010electronic}%
  \BibitemOpen
  \bibfield  {author} {\bibinfo {author} {\bibfnamefont {E.~V.}\ \bibnamefont
  {Castro}}, \bibinfo {author} {\bibfnamefont {K.~S.}\ \bibnamefont
  {Novoselov}}, \bibinfo {author} {\bibfnamefont {S.~V.}\ \bibnamefont
  {Morozov}}, \bibinfo {author} {\bibfnamefont {N.}~\bibnamefont {Peres}},
  \bibinfo {author} {\bibfnamefont {J.~L.}\ \bibnamefont {Dos~Santos}},
  \bibinfo {author} {\bibfnamefont {J.}~\bibnamefont {Nilsson}}, \bibinfo
  {author} {\bibfnamefont {F.}~\bibnamefont {Guinea}}, \bibinfo {author}
  {\bibfnamefont {A.}~\bibnamefont {Geim}}, \ and\ \bibinfo {author}
  {\bibfnamefont {A.~C.}\ \bibnamefont {Neto}},\ }\href
  {https://doi.org/10.1088/0953-8984/22/17/175503} {\bibfield  {journal}
  {\bibinfo  {journal} {J. Phys.: Condens. Matter}\ }\textbf {\bibinfo {volume}
  {22}},\ \bibinfo {pages} {175503} (\bibinfo {year} {2010})}\BibitemShut
  {NoStop}%
\bibitem [{\citenamefont {Lee}\ \emph {et~al.}(2016)\citenamefont {Lee},
  \citenamefont {Lee}, \citenamefont {Eo}, \citenamefont {Kurdak},\ and\
  \citenamefont {Zhong}}]{lee2016evidence}%
  \BibitemOpen
  \bibfield  {author} {\bibinfo {author} {\bibfnamefont {K.}~\bibnamefont
  {Lee}}, \bibinfo {author} {\bibfnamefont {S.}~\bibnamefont {Lee}}, \bibinfo
  {author} {\bibfnamefont {Y.~S.}\ \bibnamefont {Eo}}, \bibinfo {author}
  {\bibfnamefont {C.}~\bibnamefont {Kurdak}}, \ and\ \bibinfo {author}
  {\bibfnamefont {Z.}~\bibnamefont {Zhong}},\ }\href
  {https://doi.org/10.1103/PhysRevB.94.205418} {\bibfield  {journal} {\bibinfo
  {journal} {Phys. Rev. B}\ }\textbf {\bibinfo {volume} {94}},\ \bibinfo
  {pages} {205418} (\bibinfo {year} {2016})}\BibitemShut {NoStop}%
\bibitem [{\citenamefont {Mak}\ \emph {et~al.}(2009)\citenamefont {Mak},
  \citenamefont {Lui}, \citenamefont {Shan},\ and\ \citenamefont
  {Heinz}}]{mak2009observation}%
  \BibitemOpen
  \bibfield  {author} {\bibinfo {author} {\bibfnamefont {K.~F.}\ \bibnamefont
  {Mak}}, \bibinfo {author} {\bibfnamefont {C.~H.}\ \bibnamefont {Lui}},
  \bibinfo {author} {\bibfnamefont {J.}~\bibnamefont {Shan}}, \ and\ \bibinfo
  {author} {\bibfnamefont {T.~F.}\ \bibnamefont {Heinz}},\ }\href
  {https://doi.org/10.1103/PhysRevLett.102.256405} {\bibfield  {journal}
  {\bibinfo  {journal} {Phys. Rev. Lett.}\ }\textbf {\bibinfo {volume} {102}},\
  \bibinfo {pages} {256405} (\bibinfo {year} {2009})}\BibitemShut {NoStop}%
\bibitem [{\citenamefont {Alymov}\ \emph {et~al.}(2016)\citenamefont {Alymov},
  \citenamefont {Vyurkov}, \citenamefont {Ryzhii},\ and\ \citenamefont
  {Svintsov}}]{alymov2016abrupt}%
  \BibitemOpen
  \bibfield  {author} {\bibinfo {author} {\bibfnamefont {G.}~\bibnamefont
  {Alymov}}, \bibinfo {author} {\bibfnamefont {V.}~\bibnamefont {Vyurkov}},
  \bibinfo {author} {\bibfnamefont {V.}~\bibnamefont {Ryzhii}}, \ and\ \bibinfo
  {author} {\bibfnamefont {D.}~\bibnamefont {Svintsov}},\ }\href
  {https://doi.org/10.1038/srep24654} {\bibfield  {journal} {\bibinfo
  {journal} {Sci. Rep.}\ }\textbf {\bibinfo {volume} {6}},\ \bibinfo {pages}
  {24654} (\bibinfo {year} {2016})}\BibitemShut {NoStop}%
\bibitem [{\citenamefont {Skinner}\ \emph {et~al.}(2014)\citenamefont
  {Skinner}, \citenamefont {Shklovskii},\ and\ \citenamefont
  {Voloshin}}]{skinner2014bound}%
  \BibitemOpen
  \bibfield  {author} {\bibinfo {author} {\bibfnamefont {B.}~\bibnamefont
  {Skinner}}, \bibinfo {author} {\bibfnamefont {B.}~\bibnamefont {Shklovskii}},
  \ and\ \bibinfo {author} {\bibfnamefont {M.}~\bibnamefont {Voloshin}},\
  }\href {https://doi.org/10.1103/PhysRevB.89.041405} {\bibfield  {journal}
  {\bibinfo  {journal} {Phys. Rev. B}\ }\textbf {\bibinfo {volume} {89}},\
  \bibinfo {pages} {041405} (\bibinfo {year} {2014})}\BibitemShut {NoStop}%
\bibitem [{\citenamefont {Liu}\ \emph {et~al.}(2011)\citenamefont {Liu},
  \citenamefont {Feng},\ and\ \citenamefont {Yao}}]{Liu2011}%
  \BibitemOpen
  \bibfield  {author} {\bibinfo {author} {\bibfnamefont {C.-C.}\ \bibnamefont
  {Liu}}, \bibinfo {author} {\bibfnamefont {W.}~\bibnamefont {Feng}}, \ and\
  \bibinfo {author} {\bibfnamefont {Y.}~\bibnamefont {Yao}},\ }\href
  {http://link.aps.org/doi/10.1103/PhysRevLett.107.076802} {\bibfield
  {journal} {\bibinfo  {journal} {Phys. Rev. Lett.}\ }\textbf {\bibinfo
  {volume} {107}},\ \bibinfo {pages} {076802} (\bibinfo {year}
  {2011})}\BibitemShut {NoStop}%
\bibitem [{\citenamefont {Drummond}\ \emph {et~al.}(2012)\citenamefont
  {Drummond}, \citenamefont {Z\'olyomi},\ and\ \citenamefont
  {Fal'ko}}]{Drummond2012}%
  \BibitemOpen
  \bibfield  {author} {\bibinfo {author} {\bibfnamefont {N.~D.}\ \bibnamefont
  {Drummond}}, \bibinfo {author} {\bibfnamefont {V.}~\bibnamefont {Z\'olyomi}},
  \ and\ \bibinfo {author} {\bibfnamefont {V.~I.}\ \bibnamefont {Fal'ko}},\
  }\href {http://link.aps.org/doi/10.1103/PhysRevB.85.075423} {\bibfield
  {journal} {\bibinfo  {journal} {Phys. Rev. B}\ }\textbf {\bibinfo {volume}
  {85}},\ \bibinfo {pages} {075423} (\bibinfo {year} {2012})}\BibitemShut
  {NoStop}%
\bibitem [{\citenamefont {Ezawa}(2012{\natexlab{a}})}]{Ezawa_prl2012}%
  \BibitemOpen
  \bibfield  {author} {\bibinfo {author} {\bibfnamefont {M.}~\bibnamefont
  {Ezawa}},\ }\href {http://link.aps.org/doi/10.1103/PhysRevLett.109.055502}
  {\bibfield  {journal} {\bibinfo  {journal} {Phys. Rev. Lett.}\ }\textbf
  {\bibinfo {volume} {109}},\ \bibinfo {pages} {055502} (\bibinfo {year}
  {2012}{\natexlab{a}})}\BibitemShut {NoStop}%
\bibitem [{\citenamefont {Ezawa}(2013)}]{Ezawa_prl2013}%
  \BibitemOpen
  \bibfield  {author} {\bibinfo {author} {\bibfnamefont {M.}~\bibnamefont
  {Ezawa}},\ }\href {http://link.aps.org/doi/10.1103/PhysRevLett.110.026603}
  {\bibfield  {journal} {\bibinfo  {journal} {Phys. Rev. Lett.}\ }\textbf
  {\bibinfo {volume} {110}},\ \bibinfo {pages} {026603} (\bibinfo {year}
  {2013})}\BibitemShut {NoStop}%
\bibitem [{\citenamefont {Ho}\ and\ \citenamefont
  {Lan}(2016)}]{ho2016photoenhanced}%
  \BibitemOpen
  \bibfield  {author} {\bibinfo {author} {\bibfnamefont {L.~B.}\ \bibnamefont
  {Ho}}\ and\ \bibinfo {author} {\bibfnamefont {T.~N.}\ \bibnamefont {Lan}},\
  }\href {https://doi.org/10.1088/0022-3727/49/37/375106} {\bibfield  {journal}
  {\bibinfo  {journal} {J. Phys. D: Appl. Phys.}\ }\textbf {\bibinfo {volume}
  {49}},\ \bibinfo {pages} {375106} (\bibinfo {year} {2016})}\BibitemShut
  {NoStop}%
\bibitem [{\citenamefont {Ezawa}(2012{\natexlab{b}})}]{ezawa2012quasi}%
  \BibitemOpen
  \bibfield  {author} {\bibinfo {author} {\bibfnamefont {M.}~\bibnamefont
  {Ezawa}},\ }\href {https://doi.org/10.1143/JPSJ.81.104713} {\bibfield
  {journal} {\bibinfo  {journal} {Journal of the Physical Society of Japan}\
  }\textbf {\bibinfo {volume} {81}},\ \bibinfo {pages} {104713} (\bibinfo
  {year} {2012}{\natexlab{b}})}\BibitemShut {NoStop}%
\bibitem [{\citenamefont {Da}\ \emph {et~al.}(2017)\citenamefont {Da},
  \citenamefont {Ding},\ and\ \citenamefont {Yan}}]{da2017magneto}%
  \BibitemOpen
  \bibfield  {author} {\bibinfo {author} {\bibfnamefont {H.}~\bibnamefont
  {Da}}, \bibinfo {author} {\bibfnamefont {W.}~\bibnamefont {Ding}}, \ and\
  \bibinfo {author} {\bibfnamefont {X.}~\bibnamefont {Yan}},\ }\href
  {https://doi.org/10.1063/1.4979589} {\bibfield  {journal} {\bibinfo
  {journal} {Appl. Phys. Lett.}\ }\textbf {\bibinfo {volume} {110}},\ \bibinfo
  {pages} {141105} (\bibinfo {year} {2017})}\BibitemShut {NoStop}%
\bibitem [{\citenamefont {Huang}\ \emph {et~al.}(2014)\citenamefont {Huang},
  \citenamefont {Deng}, \citenamefont {Lee}, \citenamefont {Yoon},
  \citenamefont {Sumpter}, \citenamefont {Liu}, \citenamefont {Smith},\ and\
  \citenamefont {Wei}}]{huang2014exceptional}%
  \BibitemOpen
  \bibfield  {author} {\bibinfo {author} {\bibfnamefont {B.}~\bibnamefont
  {Huang}}, \bibinfo {author} {\bibfnamefont {H.-X.}\ \bibnamefont {Deng}},
  \bibinfo {author} {\bibfnamefont {H.}~\bibnamefont {Lee}}, \bibinfo {author}
  {\bibfnamefont {M.}~\bibnamefont {Yoon}}, \bibinfo {author} {\bibfnamefont
  {B.~G.}\ \bibnamefont {Sumpter}}, \bibinfo {author} {\bibfnamefont
  {F.}~\bibnamefont {Liu}}, \bibinfo {author} {\bibfnamefont {S.~C.}\
  \bibnamefont {Smith}}, \ and\ \bibinfo {author} {\bibfnamefont {S.-H.}\
  \bibnamefont {Wei}},\ }\href {https://doi.org/10.1103/PhysRevX.4.021029}
  {\bibfield  {journal} {\bibinfo  {journal} {Phys. Rev. X}\ }\textbf {\bibinfo
  {volume} {4}},\ \bibinfo {pages} {021029} (\bibinfo {year}
  {2014})}\BibitemShut {NoStop}%
\end{thebibliography}%

\end{document}